\documentclass[aps,prx,twocolumn,notitlepage,superscriptaddress]{revtex4-1}
\usepackage[colorlinks=true, citecolor=magenta, linkcolor=blue, urlcolor=blue]{hyperref}
\usepackage{graphicx}
\usepackage{amsmath}
\usepackage{amssymb}
\usepackage{amsfonts}
\usepackage{xcolor}
\usepackage{tabularx}
\usepackage{tikz}
\usepackage{comment}
\usepackage{braket}

\newcommand{\bS}{{\bf S}}
\newcommand{\br}{{\bf r}}

\newcommand{\bk}{{\bf k}}
\newcommand{\bq}{{\bf q}}

\newcommand\T{\rule{0pt}{2.6ex}}       
\newcommand\B{\rule[-1.2ex]{0pt}{0pt}} 

\allowdisplaybreaks

\begin{document}
 
\title{Cavity control of multiferroic order in single-layer NiI$_2$}

\author{Chongxiao Fan}
\thanks{These authors contributed equally to this work. \\ \href{mailto:emil.bostrom@mpsd.mpg.de}{emil.bostrom@mpsd.mpg.de}}
\affiliation{Max Planck Institute for the Structure and Dynamics of Matter, Luruper Chaussee 149, 22761 Hamburg, Germany}
\affiliation{Institute for Theory of Statistical Physics, RWTH Aachen University, and JARA Fundamentals of Future Information Technology, 52062 Aachen, Germany}

\author{Emil {Vi\~nas Bostr\"om}}
\thanks{These authors contributed equally to this work. \\ \href{mailto:emil.bostrom@mpsd.mpg.de}{emil.bostrom@mpsd.mpg.de}}
\affiliation{Max Planck Institute for the Structure and Dynamics of Matter, Luruper Chaussee 149, 22761 Hamburg, Germany}
\affiliation{Nano-Bio Spectroscopy Group and ETSF,  Departamento de Polímeros y Materiales Avanzados: Fisica, Química y Tecnologia, Universidad del País Vasco UPV/EHU- 20018 San Sebastián, Spain}

\author{Xinle Cheng}
\affiliation{Max Planck Institute for the Structure and Dynamics of Matter, Luruper Chaussee 149, 22761 Hamburg, Germany}

\author{Lukas Grunwald}
\affiliation{Max Planck Institute for the Structure and Dynamics of Matter, Luruper Chaussee 149, 22761 Hamburg, Germany}
\affiliation{Institute for Theory of Statistical Physics, RWTH Aachen University, and JARA Fundamentals of Future Information Technology, 52062 Aachen, Germany}

\author{Zhuquan Zhang}
\affiliation{Department of Physics, Columbia University, New York, NY 10027, USA}

\author{Dante M. Kennes}
\affiliation{Institute for Theory of Statistical Physics, RWTH Aachen University, and JARA Fundamentals of Future Information Technology, 52062 Aachen, Germany}
\affiliation{Max Planck Institute for the Structure and Dynamics of Matter, Luruper Chaussee 149, 22761 Hamburg, Germany}

\author{Dmitri N. Basov}
\affiliation{Department of Physics, Columbia University, New York, NY 10027, USA}

\author{Angel Rubio}
\email{angel.rubio@mpsd.mpg.de}
\affiliation{Max Planck Institute for the Structure and Dynamics of Matter, Luruper Chaussee 149, 22761 Hamburg, Germany}
\affiliation{Initiative for Computational Catalysis, Flatiron Institute, Simons Foundation, New York City, NY 10010, USA}
\affiliation{Nano-Bio Spectroscopy Group and ETSF,  Departamento de Polímeros y Materiales Avanzados: Fisica, Química y Tecnologia, Universidad del País Vasco UPV/EHU- 20018 San Sebastián, Spain}
\date{\today}

\begin{abstract}
 Controlling materials through their interactions with electromagnetic vacuum fluctuations is an emergent frontier in material engineering. Although recent experiments have demonstrated dark cavity effects for electronic material phases, like superconductivity, ferroelectricity and charge density waves, a smoking gun experiment for magnetic systems is lacking. Largely, this comes from the focus on phase transitions, where a large critical light-matter coupling is needed to observe cavity modifications.
 Here, we propose spiral magnets, where even a small cavity-mediated change in magnetic interactions is reflected in a change of the spiral wavelength, as a promising platform to observe cavity effects. We focus on the single-layer multiferroic NiI$_2$, interacting with electric field fluctuations from surface phonon polaritons of the paraelectric substrate SrTiO$_3$. With decreasing substrate-material distance, the ratio of nearest and third nearest neighbor exchange interactions reduces, leading to an increase of the spiral wavelength and an eventual transition into a ferromagnetic state. Our work identifies a realistic platform to observe cavity vacuum renormalization effects in magnetic systems.
\end{abstract}

\maketitle


\section*{Introduction}
The demonstration of magnetic order in van der Waals (vdW) materials, down to the single-layer limit, has revived interest in two-dimensional magnetism~\cite{Gong17,Burch18,Gong19}. The combination of low dimensionality, frustrated interactions, and spin-orbit coupling induced anisotropy makes vdW magnets prime candidates to explore exotic quantum phases such as valence bond order and quantum spin liquids~\cite{Savary2016,Takagi19}, supporting emergent excitations with unconventional properties. In addition, the electronic, magnetic and optical properties of these materials are sensitive to a wide range of material engineering techniques, such as strain~\cite{Kim2018,Cenker2022}, nanostructuring~\cite{Moll2018}, electric fields~\cite{Huang2018,Jiang2018}, optical pumping~\cite{Gao2024}, and moir\'e twisting~\cite{Xie2021,Kennes2021}, allowing their state to be exquisitely tuned with high precision. Recent progress has also established optical engineering techniques as a means to functionalize materials and to reach exotic out-of-equilibrium phases by strong ultrafast laser pulses~\cite{McIver2019,Rudner2020,Shan2021,delaTorre2021,Bloch2022,Ilyas2023}. However, driving a system with lasers is often associated with excessive heating and damaging, and the induced effects are intrinsically transient~\cite{DAlessio2014,Kennes2018}, thereby hindering on-chip applications.

A large effort is presently being made to use structured electromagnetic environments, such as optical cavities, dielectric interfaces or photonic crystals, to control materials via their interaction with electromagnetic vacuum fluctuations~\cite{Latini2021,Appugliese2022,VinasBostrom2023,Dirnberger2023,Jarc2023,disa_polarizing_2020,Disa2021,lu_cavity_2025,lu_cavity-enhanced_2024,tay_multimode_2025,kim_observation_2025,baydin_perspective_2025,kim_symmetry-controlled_2025}. This approach rests on the idea that changing a material's electromagnetic environment modifies the structure of the photon modes, and thereby the electromagnetic vacuum fluctuations, which in turn can alter a material's equilibrium properties~\cite{Hubener2024,lu_cavity_2025}. Recent experiments have demonstrated such dark cavity modifications of charge density wave, (fractional) quantum Hall and superconducting systems~\cite{Jarc2023,Appugliese2022,Enkner2025,Keren2025}, and suggested that the underlying physical mechanisms are highly off-resonant, and therefore sensitive to changes in the photon mode structure at all frequencies. This makes vacuum dressing of equilibrium material properties qualitatively different from polaritonic physics, where real photons are resonantly coupled to excitations of matter, and motivates the introduction of the terminology {\it endyonic} physics~\footnote{This term arises from the Greek {\it endyo} (``to put on, to be covered, to clothe/dress''), and is used to denote matter degrees of freedom dressed by vacuum fluctuations. It was coined during the Flagship School on ``Ab Initio Quantum Electrodynamics for Quantum Materials Engineering'' held at the Flatiron Institute in New York City, September 29–October 3, 2025.}. To accurately describe such effects it is necessary to develop a theoretical framework that accounts for the full mode structure of the electromagnetic field, as well as the complexities of the material. 

While recent experiments have demonstrated cavity control over electronic phases, a smoking gun experiment is still lacking for magnetic systems. To a large extent this comes from the focus on phase transitions (e.g., between antiferromagnetic and ferromagnetic states) instead of gradual material changes, where a large light-matter coupling is required for cavity modifications to be observable. In this context, spiral magnets appear as a promising platform to observe cavity effects, since in such systems even a small cavity-mediated change in magnetic interactions is directly reflected in a change of the spiral wavelength.

\begin{figure*}
\includegraphics[width=\linewidth]{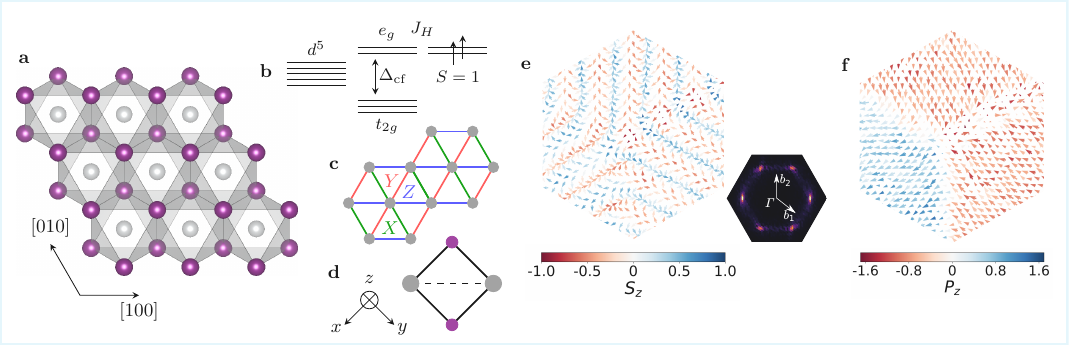}
 \caption{{\bf Magnetic and electric properties of NiI$_2$.} {\bf a,} Crystal structure of single layer NiI$_2$. The gray Ni ions form a triangular lattice, and are each surrounded by an octahedron of purple I ions. {\bf b,} Schematic illustration of the local Ni electronic states, underlying the formation of local $S = 1$ moments. The crystal field splits the $d$-orbitals into a fully occupied $t_{2g}$ manifold and an $e_g$-manifold with two electrons. Local Hund's coupling favors spin alignment between these electrons, leading to the formation of an $S = 1$ magnetic moment. {\bf c,} Division of nearest neighbor bonds into $X$-, $Y$- and $Z$-types. {\bf d,} Nearest neighbor Ni-Ni cluster on a $Z$-bond, used for the numerical down-folding. {\bf e,} Real space magnetic structure calculated from the spin Hamiltonian of Eq.~\ref{eq:spin_ham}, assuming classical spins of length $S = 1$. The arrows show the in-plane direction of the spins, and their color the out-of-plane component. The structure illustrates three magnetic domains with momenta ${\bf q}$ related by $120^\circ$ rotation, giving the momentum space spin structure factor shown in the inset. {\bf f,} Polarization distribution corresponding to the magnetic structure in {\bf e}, calculated from the bond polarization in Eq.~\ref{eq:bond_polarization}.}
 \label{fig:structure}
\end{figure*}

Here, we show how the intertwined electric and magnetic orders of the two-dimensional multiferroic NiI$_2$ can be controlled by interfacing the material with a polaritonic surface cavity. Starting from a microscopic model of NiI$_2$, we show how the magnetic interactions are modified by the coupling to surface phonon polariton fluctuations from the paraelectric substrate SrTiO$_3$. We find that as a function of substrate-material distance, the ratio $|J_1|/J_3$ of the nearest and third nearest neighbor exchange interactions is increased, leading to an elongation of the spiral wavelength and the eventual transition into a ferromagnetic state. Our work thereby identifies a realistic platform to observe cavity renormalization effects in magnetic systems and demonstrate the ideas of cavity engineering of magnetic phases.


\section*{Results}

\subsection*{Microscopic electronic model}
NiI$_2$ is a van der Waals material crystallizing in a rhombohedral structure with the space group $R\bar{3}m$. It consists of layers of Ni ions forming a triangular lattice, surrounded by edge-sharing I octahedra (see Fig.~\ref{fig:structure}a). Recent experiments have found that bulk (more than 10 layers) NiI$_2$ orders first into a collinear antiferromagnet below $T_N \approx 75$~K, and then into a helical state below $T_N \approx 60$ K~\cite{Song2022,Miao2025,Gao2024,Song2025,Tseng2025}, whose inter-layer propagation is close to antiferromagnetic. The critical temperature $T_N$ of the helical state decreases as the layer number decreases~\cite{ju_possible_2021}, and the single layer shows a transition into a helical state below $T_N \approx 20$ K~\cite{Song2022,Miao2025}. The spiral state spontaneously breaks inversion symmetry and allows for a ferroelectric polarization, making the system a type-II multiferroic.

To describe the equilibrium state of NiI$_2$, we construct a microscopic electronic model. As illustrated in Fig.~\ref{fig:structure}b, the octahedral crystal field splits the Ni $d$-orbitals into a lower $t_{2g}$ and a higher $e_g$ manifold. The $t_{2g}$ orbitals are completely filled in equilibrium, while the $e_g$ manifold contains two electrons bound into an $S = 1$ magnetic moment by the local Hund's coupling $J_H$. The I ions are described in terms of their $p$-orbitals, which are separated from the Ni $e_g$-orbitals by a charge-transfer energy $\Delta$ and completely filled in the ground state. The local atomic structure of both the Ni and I ions can be described by a Hubbard-Kanamori Hamiltonian~\cite{Stavropoulos2019,VinasBostrom2023}, which includes an intra-orbital Hubbard interaction $U$, an inter-orbital interaction $U' = U - 2J_H$, the Hund's coupling $J_H$, and a spin-orbit coupling (SOC) $\lambda$.

The magnetic interactions arise from a superexchange mechanism, and are determined by the competition between local interactions and virtual hopping processes described by a kinetic Hamiltonian $H_t$. Assuming $C_3$ symmetry around the $z$-axis, as well as inversion symmetry, it is sufficient to calculate the magnetic interactions explicitly on a single Ni-Ni bond, after which the interactions on the remaining bonds can be obtained by symmetry transformations~\cite{Maksimov2019}. As in honeycomb Kitaev systems~\cite{Winter2016}, the bonds can be divided into $X$-, $Y$- and $Z$-types (see Fig.~\ref{fig:structure}c), where each bond involves hopping processes along the complementary local axes.

Due to the structure of NiI$_2$, both direct Ni-Ni as well as indirect Ni-I and I-I superexchange paths contribute to the magnetic interactions and dictate the specific magnetic order at a given temperature. For the nearest neighbor interactions the indirect Ni-I processes were found to dominate, while for third nearest neighbor interactions the main contribution comes from direct processes. In fact, the third nearest neighbor exchange is of the same magnitude as the nearest neighbor exchange, indicating that orbital hybridization is a prominent feature of NiI$_2$. This is confirmed by the delocalized nature of the maximally localized Wannier functions (see Methods), which are closer to molecular than atomic orbitals. In contrast, the second nearest neighbor interaction remains small due to the lack of efficient hopping paths connecting such atoms. All electronic parameters needed to derive the magnetic interaction have been calculated from first principles, as is further discussed in the Methods.


\begin{table}[]
 \centering
 \begin{tabular}{c|c|c|c|c|c|c|c} \hline\hline
        & $J_1$ & $K$  & $\Gamma$ & $\Gamma'$ & $J_3$ & $A_{zz}$ & $B_I$ \T\B \\ \hline
    meV & -4.24 & 1.06 & 0.02 & 0.06 & 2.54 & 0.12 & -0.61 \T\B \\ \hline\hline
 \end{tabular}
 \caption{Equilibrium magnetic interaction parameters of the spin Hamiltonian in Eq.~\ref{eq:spin_ham}, calculated from the microscopic electronic model via a fourth order strong coupling expansion~\cite{VinasBostrom2023,Gao2024}.}
 \label{tab:spin_parameters}
\end{table}

\subsection*{Microscopic spin model}
From the electronic model, we derived an effective spin Hamiltonian for the Ni magnetic moments. The most general symmetry allowed spin Hamiltonian on the triangular lattice is~\cite{Maksimov2019}
\begin{align}\label{eq:spin_ham}
 H_s &= \sum_{\langle ij\rangle \gamma} {\bf S}_i 
 \begin{pmatrix} J_1 & \Gamma & \Gamma' \\ \Gamma & J_1 & \Gamma' \\ \Gamma' & \Gamma' & J_1 + K \end{pmatrix}_{\hspace{-0.15cm}\gamma=Z} \hspace*{-0.45cm}
 {\bf S}_j
 + J_3 \sum_{\langle\!\langle ij\rangle\!\rangle} \bS_i \cdot \bS_j \\ \nonumber
 &+ \sum_{\langle ij\rangle} B_I ({\bf S}_i \cdot {\bf S}_j)^2 + \sum_i A_{zz} ({\bf S}_i \cdot {\bf z})^2
\end{align}
were $J_1$ and $J_3$ are nearest and third nearest neighbor Heisenberg exchanges, $K$ is the Kitaev interaction, and $\Gamma$ and $\Gamma'$ are sub-dominant magnetic anisotropies. The index $\gamma$ denotes the type of Kitaev bonds, and here we illustrate the interaction matrix for the $Z-$type bond. For $X$- and $Y$-type bonds, the matrices are obtained by a rotation of the local spin axes by $\theta = \pm 2\pi/3$, respectively. The magnetic parameters depend on the electronic parameters in a complex manner, and were obtained numerically using a strong coupling expansion~\cite{VinasBostrom2023}, which perturbatively decouples the low- and high-energy sectors of the electronic system. For the nearest neighbor interactions, this calculation was performed up to fourth order in $H_t$, on an electronic cluster with two Ni and two I ions (see Fig.~\ref{fig:structure}d). With the calculated electronic parameters, we find the spin parameters $J_1 = -4.24$ meV, $K = 1.06$ meV, $\Gamma = 0.02$ meV and $\Gamma' = 0.06$ meV, whose values compare well with previous calculations based on the four-state method~\cite{Li2023,Gao2024}. 

For the third nearest neighbor interaction, we performed the strong coupling expansion to fourth order on an effective cluster (further discussed in the Methods), including only the highest orbital of each I ion (the others are split off by an energy $3\lambda/2$ coming from the spin-orbit coupling), and an effective hopping between third nearest neighbor Ni ions. The calculated value $J_3 = 2.54$ meV is in good agreement with earlier work~\cite{Li2023}. The magnetic interactions are summarized in Tab.~\ref{tab:spin_parameters}, where as our down-folding procedure does not allow to calculate the single-ion anisotropy $A_{zz}$ and biquadratic exchange $B_I$ directly, we fix their ratio with $J_1$ to that reported in Ref.~\cite{Li2023}.


\subsection*{Electric polarization model}
Due to the non-collinear magnetic structure, NiI$_2$ develops a finite electric polarization~\cite{Song2022,Gao2024}, whose relation to the underlying spin structure can be extracted from the down-folding procedure. Since the polarization is odd under inversion, its representation in terms of spin operators is also odd. In contrast, all terms entering the spin Hamiltonian of Eq.~\ref{eq:spin_ham} are even under inversion. For the Ni-Ni bond shown in Fig.~\ref{fig:structure}d, the polarization is restricted by symmetry to lie in the Ni-I-Ni plane, and perpendicular to the Ni-Ni bond vector~\cite{Gao2024}. Specifically, on a $Z$-bond the polarization has the form
\begin{align}\label{eq:bond_polarization}
 \hat{\bf P}_{s,ij} &= P (\hat{\bf x} + \hat{\bf y}) [\hat{\bf z} \cdot (\bS_i \times \bS_j)].
\end{align}
From the same numerical procedure used to derive the spin Hamiltonian, the components of the polarization in the local crystal axes are $P_z = 0$ and $P_x = P_y = P$, with $P = 4.1$ $\mu_B/c$. This compares well with the value $P \approx 10$ $\mu_B/c$ reported in Ref.~\cite{Gao2024}.


\subsection*{Magnetic and electric properties}
To obtain its multiferroic properties, we performed classical Monte Carlo simulations of single-layer NiI$_2$ with the parameters reported in Tab.~\ref{tab:spin_parameters}. The ground state spin texture has a threefold degeneracy corresponding to three helix propagation directions ${\bf q}_i$ separated by 120$^\circ$, as expected from the $C_3$ symmetry of the spin Hamiltonian. This makes the system highly prone to domain formation, as illustrated in Fig.~\ref{fig:structure}e. Our calculations find that for single-layer NiI$_2$, ${\bf q}$ is parallel to the $[100]$ direction and the wavelength is $L = 2\pi/{\bf |q|}\approx 5a$, consistent with experimental findings~\cite{Song2022,Amini2024}. 
We also compute the discrete Fourier transform of the real-space spin texture, to obtain the reciprocal space representation ${\bf S}_\bk$. The spectrum $|{\bf S_k}|$ shows peaks at ${\bf k} = {\bf q}$ (see inset of Fig.~\ref{fig:structure}e), whose relative amplitudes are proportional to sizes of the respective domains. 

Using the electric polarization model in Eq.~\ref{eq:bond_polarization}, we calculate the magnetically induced electric polarization of the ground state (see Fig.~\ref{fig:structure}f). In each domain, the polarization is perpendicular to the propagation vector ${\bf q}$, and its sign is determined by the spin helicity, defined as a left- or right-handed rotation along ${\bf q}$. This is consistent with the findings in Ref.~\cite{Song2022,Gao2024,Song2025}. Further, we observe an out-of-plane polarization component, which arises from a breaking of the in-plane $C_2$ rotational symmetry when the helix propagation direction is along a nearest neighbor Ni-Ni bond. For both the magnetic and electric properties, we find a good agreement between our results and reported experimental data.


\begin{figure*}
\includegraphics[width=\linewidth]{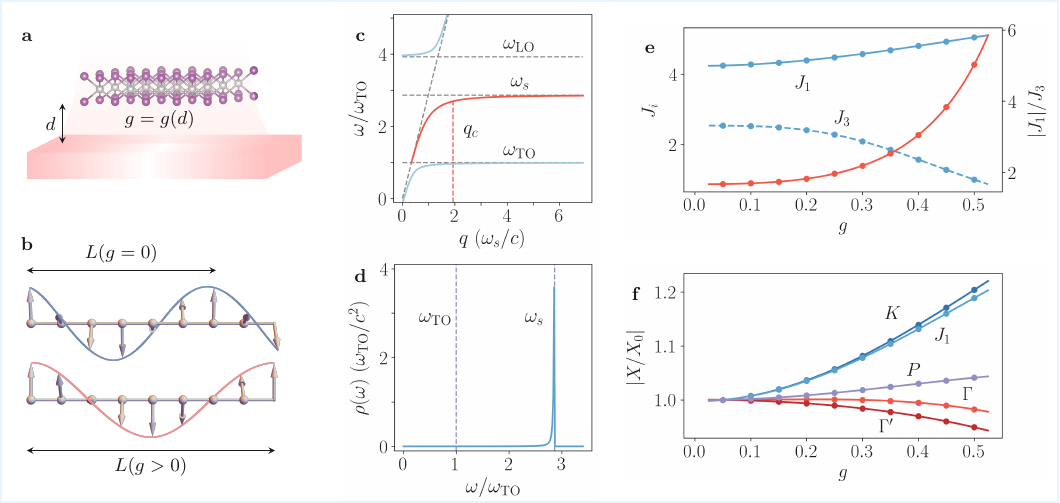}
 \caption{{\bf Principle of cavity-induced modifications of spiral magnetic order.} {\bf a,} NiI$_2$ sample deposited a distance $d$ above a paraelectric surface, generating an effective light-matter coupling $g(d)$. {\bf b,} Effect of cavity vacuum fluctuations on the magnetic structure of NiI$_2$. A finite light-matter coupling leads to an enhancement of the ratio $|J_1|/J_3$ and an elongation of the spiral wavelength. {\bf c,} Surface phonon polariton (SPP) dispersion (red) as a function of momentum. {\bf d,} SPP density of states as a function of frequency, evaluated for the parameters $\omega_s/2\pi = 2.9$ THz, $\omega_{\rm TO}/2\pi = 1.3$ THz, and $\epsilon_r = 7.3$. {\bf e,} Cavity-mediated change of the magnetic parameters $|J_1|$ (blue solid line) and $J_3$ (blue dashed line), and their ratio $|J_1|/J_3$ (red line), as a function of light-matter coupling $g$. {\bf f,} Cavity-mediated change of all nearest neighbor magnetic parameters (see Eq.~\ref{eq:spin_ham}), as well as the magnitude $P$ of the polarization operator. Dots show the results of the numerical down-folding, while lines are fits to the analytical result $X = X_0 \exp(-\alpha |g|^2) (1 + \beta |g|^2)$, showing an excellent agreement.}
 \label{fig:phase_diagram}
\end{figure*}


\subsection*{Light-matter coupling}
Having demonstrated that our model quantitatively captures the essential features of NiI$_2$, we now investigate the influence of electromagnetic vacuum fluctuations on its magnetic and electric properties. As shown below, the light-matter interaction (see Fig.~\ref{fig:phase_diagram}a) leads to an increase in the ratio $|J_1|/J_3$, and a subsequent elongation of the spiral wavelength (see Fig.~\ref{fig:phase_diagram}b). The Hamiltonian of the free photonic field is $H_{\rm phot} = \sum_{\lambda} \hbar\omega_\lambda \hat{n}_\lambda$, where $\omega_\lambda$ is the frequency of cavity mode $\lambda$, and $\hat{n}_\lambda$ is the corresponding number operator. The coupling to an electromagnetic vector potential is described via the Peierls' substitution $\hat{c}_{i\alpha\sigma}^\dagger \hat{c}_{i\beta\sigma} \to e^{i \phi_{ij}} \hat{c}_{i\alpha\sigma}^\dagger \hat{c}_{i\beta\sigma}$, where $\phi_{ij} = (ea/\hbar)\, {\bf r}_{ij} \cdot \hat{\bf A}$ is the Peierls phase, and ${\bf r}_{ij} = {\bf r}_j - {\bf r}_i$ is the vector between two ions at positions $\br_i$ and $\br_j$, measured in units of the Ni-Ni distance $a$. The quantized vector potential is $\hat{\bf A} = \sum_{\lambda} ( A_\lambda {\bf e}_\lambda \hat{a}_\lambda^\dagger + A_\lambda^{*} {\bf e}^{*}_\lambda \hat{a}_\lambda)$, and for a given mode $\lambda$ the dimensionless light-matter coupling is defined as $g_\lambda = (ea/\hbar) A_\lambda$.

Here we consider a cavity where the electric field originates from surface phonon polariton (SPP) modes of the paraelectric material SrTiO$_3$ (STO), as illustrated in Fig.~\ref{fig:phase_diagram}a. We work in the deep sub-wavelength regime $q \gg \omega/c$, where the modes are exponentially confined in the out-of-plane ($z$) direction, and primarily longitudinal. It is therefore sufficient to label the modes by a single mode index $\lambda = \bq$, the in-plane momentum. We neglect static Coulomb screening coming from the substrate~\cite{GrunwaldCheng}, which is not expected to have a large influence on the ratio of magnetic interactions. Assuming that the single-layer NiI$_2$ is a distance $d$ above the substrate, the effective light-matter coupling is~\cite{VinasBostrom2024}
\begin{align}\label{eq:light_matter_coupling}
 g_\bq &= ea e^{-qd} \sqrt{\frac{q (\omega_s^2 - \omega_{\rm TO}^2)}{4\hbar \epsilon_0 \epsilon_r \omega_s^3A}}.
\end{align}
Here, $\omega_s$ is the SPP frequency for $q \to \infty$, $\omega_{\rm TO}$ is the transverse optical phonon frequency, and $\epsilon_r = \epsilon_{\rm sub} + \epsilon_{\rm mat}$ with $\epsilon_{\rm sub}$ and $\epsilon_{\rm mat}$ being the relative permittivity of the substrate and the material, respectively. As the surface area $A$ of the cavity is large, the single mode coupling will approach zero in the macroscopic limit.


The total Hamiltonian can be expanded in the photon number basis $|{\bf n}\rangle = |n_{\bq_1}, n_{\bq_2}, \ldots, n_{\bq_N} \rangle$, and due to the structure of the strong coupling expansion, the down-folding can be performed in each photon sector separately. The result is the Hamiltonian 
\begin{align}\label{eq:spin_photon_ham}
 \mathcal{H} = \sum_{\bf nm} \Big( \mathcal{H}_{s,\bf nm} + \delta_{\bf nm} \sum_\bq \hbar\omega_\bq n_\bq \Big) |{\bf n}\rangle\langle{\bf m}|,
\end{align}
where $\mathcal{H}_{s,\bf nm}$ has the same form as in Eq.~\ref{eq:spin_ham}, but with parameters depending on the photon number sector, the cavity frequency and the light-matter coupling~\cite{VinasBostrom2023}. The form of the Hamiltonian is independent of the nature of cavity, and would look the same for a Fabry-P\'erot cavity.


\subsection*{Single effective mode approximation}
The Hamiltonian of Eq.~\ref{eq:spin_photon_ham} formally contains a macroscopic number of modes. Therefore, apart from exceptional cases where the strong coupling expansion can be performed analytically~\cite{GrunwaldCheng}, including more than a few modes in the calculation of the spin parameters is a prohibitive task. To proceed with the calculation, we therefore develop a single effective mode approximation for surface cavities (for Fabry-P\'erot cavities, see Ref.~\cite{KamperSvendsen2023}), which exactly reproduces the full multi-mode calculation for a single band Hubbard model.

The single mode approximation is motivated by noting that to leading order, vacuum modifications to the magnetic interactions are proportional to the local electric field fluctuations $\langle {\bf E}^2\rangle$, which for SPPs have the form $\epsilon_0 \langle {\bf E}^2 \rangle = \hbar^2(\omega_s^2 -\omega_{\rm TO}^2)/(16 \pi\hbar\omega_s \epsilon_r d^3)$~\cite{VinasBostrom2024}. Since the SPP density of states is highly localized around a single frequency $\omega_s$ (see Fig.~\ref{fig:phase_diagram}c and \ref{fig:phase_diagram}d), the fluctuations can be reproduced by an effective mode at $\omega = \omega_s$. The electric field for the effective mode is $\hat{\bf E} = -iE_0 \sum_\nu \hat{\bf e}_\nu ( \hat{a}_\nu^\dagger - \hat{a}_\nu )$, where $\nu$ runs over polarization states (two modes are needed to maintain the in-plane rotational symmetry), and imposing $E_0^2 = \langle {\bf E}^2\rangle/2$ results in
\begin{align}
 E_0 = \sqrt{\frac{\hbar^2(\omega_s^2 -\omega_{\rm TO}^2)}{32 \pi\hbar\omega_s \epsilon_0\epsilon_r d^3}}.
\end{align}
The effective light-matter coupling with Peierls substitution is then $g = eaE_0/(\hbar\omega_s)$. Within the single mode approximation, the light-matter coupling becomes a function of the substrate-material distance, $g \sim d^{-3/2}$. This can reach the strong coupling regime $g \approx 1$ at a distance $d = 1$ nm away from the STO surface~\cite{VinasBostrom2024}.


\begin{figure*}
\includegraphics[width=\linewidth]{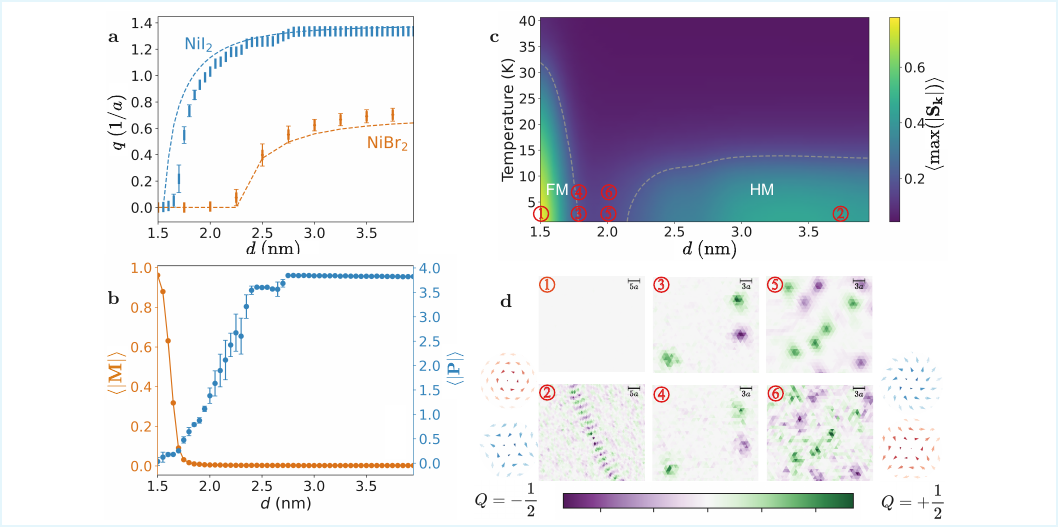}
 \caption{{\bf Cavity effects on the magnetic and electric properties of NiI$_2$.} {\bf a,} Magnitude of the spiral momentum $q = |{\bf q}|$, at zero temperature, for NiI$_2$ (blue) and NiBr$_2$ (orange). Results were obtained from spin Monte Carlo simulations of a system with $80 \times 80$ lattice sites. The bar binning reflects the finite size effects on the discrete Fourier transform. Near the critical distance $d = d_c$, the standard deviation is large because of the small energy difference between the helimagnetic and ferromagnetic states. The dashed lines show the analytical result for $q$ derived from a $J_1-J_3-A_{zz}$ model in the macroscopic limit. {\bf b,} Macroscopic magnetization $|\mathbf{M}|$ and spin-induced polarization $|\mathbf{P}|$ at zero temperature, obtained from spin Monte Carlo simulations of a system with $80 \times 80$ lattice sites. Near the transition between the helical and meron gas phases ($d \approx 2.3$ nm), the polarization exhibits a large standard deviation due to domain formation and finite size effects. {\bf c,} Magnetic phase diagram as a function of substrate-material distance $d$ and temperature $T$. The color reflects the maximal value of the spin components, $\max(|\mathbf{S_k}|)$. The dashed lines indicate the isosurface $|{\bf S(q)}|= 0.2$. {\bf d,} Calculated topological charge density for spin textures at the parameter values indicated by \textcircled{1}-\textcircled{6} in the phase diagram. In the ferromagnetic phase \textcircled{1}, there are no topological charges. In the region between the ferromagnetic and helimagnetic phases, \textcircled{3} and \textcircled{5}, isolated meron (antimeron) pairs emerge~\cite{augustin_properties_2021}. Representative spin textures corresponding to the purple regions and green regions are shown to the left (negative charge) and to the right (positive charge) of the charge densities, respectively. As the distance $d$ increases the meron density increases, and in the helimagnetic phase \textcircled{2} get bound into meron-antimeron chains at the domain walls between different helical domains. As the temperature is increased, the merons broaden (\textcircled{4} and \textcircled{6}) and disappear around $T \approx 8$ K. All numerical results were obtained from classical spin Monte Carlo simulations of the Hamiltonian in Eq.~\ref{eq:spin_photon_ham} in the dark cavity limit $|{\bf n}\rangle = |{\bf m}\rangle = |0\rangle$, using the parameters of Fig.~\ref{fig:phase_diagram}.}
 \label{fig:numerics}
\end{figure*}


\subsection*{Benchmarking the single mode approximation}
To benchmark the single-mode approximation, we consider cavity modifications within a simplified model. While it is generally prohibitive to incorporate the full cavity mode structure when calculating magnetic interactions, this is possible for a model where the isotropic interactions $J_1$ and $J_3$ are derived from an effective Ni--Ni hopping between the states of the $e_g$-manifold (see Supplemental Material for details). The result is the effective parameters $J_1 = -t_1^2/(U+J_H)$ and $J_3 = t_3^2/(U+J_H)$, where $t_1$ and $t_3$ are chosen to reproduce the parameters of the more complex model.

Using the simplified model, it is possible to calculate the cavity modification including all modes and to all orders in the light-matter coupling~\cite{GrunwaldCheng}. The final result is proportional to the photonic density of states (DOS) difference $\rho - \rho_0$, where $\rho$ and $\rho_0$ are the DOS of the SPP modes and of free space, respectively. Noting that the local DOS of the surface modes is highly localized around the frequency $\omega = \omega_s$, we approximate $\rho(\omega) - \rho_0(\omega) = r \delta(\omega - \omega_s)$, where $r$ is a fitting parameter. The expression for $J_i$ (with $i \in \{1,3\}$), in the zero photon sector, can then be written as 
\begin{align}\label{eq:resummation_interaction}
 \frac{J_i}{J_i^0} &= e^{-\alpha_i} \int_0^\infty dx\, \exp \Big[ \alpha_i e^{-\gamma x} - x \Big],
\end{align}
where $\alpha_i = e^2a_i^2 r/(2\hbar\epsilon_0\epsilon_r\omega_s)$ and $\gamma = \omega_s/U$. The parameter $r$ can be fixed by expanding the exact result to leading order in $\alpha$, and comparing to a perturbative calculation (see Methods for details)~\cite{GrunwaldCheng}. This results in $r = 1/(16\pi d^3)$, such that $\alpha_i = |g_i|^2$ is the light-matter coupling for a bond of length $a_i$. When the integral above is evaluated, it exactly agrees with the result from the single mode approximation.


\subsection*{Cavity renormalized magnetic interactions}
As the single mode approximation and multi-mode calculation exactly agree on the cavity-induced exchange modification of a single-band Hubbard model, we can use it to evaluate the modification of magnetic interactions within the more complex model. This provides access to all nearest neighbor magnetic interactions, and to the electric polarization. 
To gain some intuition, it is however useful to first consider the simplified model, where the single-mode expression for the cavity renormalization of the magnetic parameters is
\begin{align}
 \frac{X}{X_0} &= e^{-|g|^2} \sum_n |g|^{2n} \frac{\Omega}{\Omega + \omega_s n}.
\end{align}
Apart from natural constants, the renormalization depends only on the substrate-material distance $d$, the cavity frequency $\omega_s$, and the effective bond length $a$. Using the values $\omega_s = 5$ THz, a relative permittivity $\epsilon_r = 4$, and a distance $d = 10$ nm, the coupling parameter is $|g_i|^2 \approx \alpha_0 (a_i^2/d^2)$ with $\alpha_0 = 1.37$. Keeping only the first term in the sum, we have $X/X_0 = \exp(-|g|^2)$, and the ratio of the nearest and third nearest neighbor exchanges is
\begin{align}
 \frac{J_1}{J_3} &= \frac{J_{1}^0}{J_{3}^0} \exp\Big( \alpha_0 \frac{a_3^2 - a_1^2}{d^2} \Big).
\end{align}
While cavity fluctuations suppress both $J_1$ and $J_3$, the effect is proportional to the bond length $a_i$, which is two times larger for $J_3$. The ratio $J_1/J_3$ therefore increases as the material gets closer to the paraelectric surface.

For the numerical results, only a finite number of photon states can be included in the calculation. The numerical results are therefore compared to a series expansion of Eq.~\ref{eq:resummation_interaction} in $|g|^2$, where $X/X_0 = \exp(-\alpha|g|^2) (1 + \beta|g|^2)$. As shown in Fig.~\ref{fig:phase_diagram}e and \ref{fig:phase_diagram}f, this function provides a very good fit both to the isotropic interactions $J_1$ and $J_3$, and to the highly anisotropic interactions $K$, $\Gamma$ and $\Gamma'$. We note that this is a non-trivial result, as the form of the renormalization was derived within the simplified electronic model, while the numerical results are obtained from a complex multi-band model. The excellent agreement between numerical and analytical results validates the consistency of the present framework, and enables an efficient interpolation of the magnetic interactions to arbitrary light-matter couplings.


\subsection*{Cavity modification of magnetic order}
Using the magnetic interactions of Figs.~\ref{fig:phase_diagram}e and \ref{fig:phase_diagram}f, we now calculate the equilibrium magnetic state as a function of the substrate-material distance $d$. For this we consider the classical limit of the Hamiltonian in Eq.~\ref{eq:spin_photon_ham} in the dark cavity limit $|{\bf n}\rangle = |{\bf m}\rangle = |0\rangle$, where all interactions are functions of the distance dependent light-matter coupling $g(d)$, and the spins are classical vectors of length $|{\bf S}| = 1$. The magnetic ground state is found by spin Monte Carlo simulations.

For $d \to \infty$, the calculated spin parameters reproduce the experimentally observed ground state, corresponding to a helical magnetic state with wavelength $L \approx 5a$ propagating along the $[100]$ direction. As the distance is decreased, $J_3$ is suppressed relative to $J_1$, resulting in an elongation of the spin spiral (a decrease in $|{\bf q}|$) as shown in Fig.~\ref{fig:numerics}a. As the distance falls below a critical value $d_c \approx 1.6$ nm, the system fully transitions to a ferromagnetic state. This is signaled by a build-up of the magnetization $\langle|{\bf M}|\rangle$, and a vanishing electric polarization $\langle|{\bf P}|\rangle$ (see Fig.~\ref{fig:numerics}b).

To supplement the Monte Carlo simulations, and rigorously address the macroscopic limit, we analytically solve the classical spin model for $K = \Gamma = \Gamma' = 0$ (see Supplemental Material). This is justified by noting that the equilibrium magnetic order is dominated by the competition between a ferromagnetic $J_1$ and an antiferromagnetic $J_3$. The analytical solution predicts a transition from a helical state with $\bq \parallel [100]$ to a ferromagnetic state at a ratio $|J_1|/J_3 = 4$, and as a function of couplings the spiral momentum is $q = 2\arccos[\frac{1}{4}(1 + \sqrt{1 + 2|J_1|/J_3})]$. The transition happens at a critical distance $d_c \approx 1.6$ nm, in very good agreement with the Monte Carlo simulations of NiI$_2$.

We also map out the full magnetic phase diagram as a function of distance $d$ and temperature $T$, using the Fourier magnitude $|{\bf S(q)}|$ as the order parameter. The result is shown in Fig.~\ref{fig:numerics}c, where the contour lines delineate three identified regions of the low temperature magnetic state. Contrary to the prediction of the analytical solution, we find that the phase transition from the helimagnetic to the ferromagnetic state is not sharp. Instead, our simulations reveal a crossover region between the two phases, where local spin vortices and anti-vortices appear (see Fig.~\ref{fig:numerics}d, panels 3 and 5). Further calculations identify the vortices as merons and anti-merons~\cite{augustin_properties_2021,bachmann_meron_2023} with a topological charge $Q = \pm 1/2$. The highest meron density is found close to the spiral phase, and decreases towards to ferromagnetic state. While chains of merons also appear at the domain walls of the helical phase~\cite{wang_microscopic_2025} (see Fig.~\ref{fig:numerics}d, panel 2), they become unbound in the intermediate regime and form a gas-like state. In the gas-like state the merons broaden with temperature (Fig.~\ref{fig:numerics}d, panel 4 and 6), but persist up to $T \approx 8$ K.


\section*{Discussion}
By developing a first-principles based microscopic model for NiI$_2$, coupled to a quantum electromagnetic field, we have demonstrated that the ratio $|J_1|/J_3$ determining the equilibrium magnetic order is sensitive to changes in cavity vacuum fluctuations. This result is consistently predicted both by numerical calculations based on a complex microscopic electronic model, and by analytical calculations on a simplified model. In both cases, the ratio $|J_1|/J_3$ increases exponentially with the light-matter coupling $|g|^2$, which varies as $|g|^2 \sim d^{-3}$ with the substrate-material distance $d$. The modification of the magnetic interactions is reflected in a change of the equilibrium magnetic state, where the spiral wavelength $L$ (spiral momentum $q = |{\bf q}|$) gradually increases (decreases) as $d$ is made smaller, until the system turns ferromagnetic below $d_c \approx 1.6$ nm. Between the helical and ferromagnetic phases, we further identify a crossover region with a large classical ground state degeneracy. A further analysis shows that the spin stiffness goes to zero at the phase boundary, even as $T \to 0$ (see Supplemental Material), indicating the presence of a quantum critical point in the intermediate region that would motivate further study.

The cavity mediated continuous change of the equilibrium magnetic order is reflected in a number of observable properties. The transition from a helimagnetic to a ferromagnetic state is associated with a finite magnetization, and should be possible to probe by magneto-optical Kerr or Faraday rotation measurements~\cite{Gao2024}. Concurrently, the transition into a ferromagnetic state leads to the restoration of inversion symmetry, and a vanishing of the ferroelectric polarization, which should be reflected in a sharp decrease in second harmonic generation and linear dichroism~\cite{Song2022,Gao2024}. Moving beyond far-field optical signals, the cavity fluctuations lead to a gradual change in the spiral wavelength and momentum, which could be probed either with neutron diffraction, resonant X-ray scattering, scanning tunneling or atomic force microscopy~\cite{kuindersma1981magnetic,Tseng2025,Miao2025,Wang2024,Amini2024,Amini2025}, or possibly also with scanning near-field optical microscopy. We also note that the iso-structural material NiBr$_2$ has a longer spiral wavelength, and therefore a larger original value of the ratio $|J_1|/J_3$. For this reason, the helimagnetic to ferromagnetic transition occurs at a larger distance $d_c$ in NiBr$_2$, as shown in Fig.~\ref{fig:numerics}a.

In contrast to most earlier proposed modifications of material properties by cavity vacuum fluctuations, the cavity-mediated magnetic changes here happen gradually with the substrate-material distance. Therefore, it is not necessary to reach a critical effective light-matter coupling to observe the effects. We note that in the few cases where effects of cavity vacuum fluctuations on material properties have been observered~\cite{Appugliese2022,Jarc2023,Enkner2025,Keren2025}, this has been a key requirement to resolve them. Our work thereby presents a realistic proposal for a smoking gun experimental verification of cavity modified magnetism.

\section*{Acknowledgements and funding}
We acknowledge support from the Cluster of Excellence “CUI: Advanced Imaging of Matter”–EXC 2056–project ID 390715994, SFB-925 “Light induced dynamics and control of correlated quantum systems”–project ID 170620586 of the Deutsche Forschungsgemeinschaft (DFG), the European Research Council (ERC-2024-SyG-UnMySt–101167294), and the Max Planck-New York City Center for Non-Equilibrium Quantum Phenomena. D.M.K. acknowledges support by the DFG via Germany's Excellence Strategy-Cluster of Excellence Matter and Light for Quantum Computing (ML4Q, Project No. EXC 2004/1, Grant No.390534769), individual Grant No. 508440990 and 531215165 (Research Unit ``OPTIMAL''). The Flatiron Institute is a division of the Simons Foundation.

\section*{Author contributions} The work was conceptualized by EVB and AR. The methodology was developed by CF, EVB, XC and LG, and the main investigations performed by CF and EVB. ZZ and DNB provided expertise on how to connect the theoretical predictions to experimental observables. The work was supervised by AR and DMK, and all authors contributed to the writing and editing of the final manuscript.

\section*{Competing interests} The authors declare that they have no competing interests. 

\section*{Data and materials availability} All data needed to evaluate the conclusions in the paper are present in the paper and/or the Supplementary Materials.


\clearpage

\section*{Methods}

\subsection*{Local electronic Hamiltonian}
We here present the microscopic electronic model used to calculate cavity-modified magnetic interactions. Due to the crystal field, the Ni $d$-orbitals are split into a lower $t_{2g}$ and a higher $e_g$ manifold, the latter consisting of the $d_{x^2-y^2}$ and $d_{3z^2-r^2}$ orbitals. The $t_{2g}$ orbitals are assumed to be completely filled, while the $e_g$ orbitals contain two electrons forming an effective $S = 1$ magnetic moment. The I $p$-orbitals ($p_x$, $p_y$ and $p_z$) are separated from the Ni $e_g$-orbitals by a charge-transfer energy $\Delta$, and are assumed to be completely filled in the ground state. The local properties of both Ni and I ions are described by the Hubbard-Kanamori Hamiltonian
\begin{align}\label{eq:kanamori}
 H_U &= U \sum_{\alpha} \hat{n}_{\alpha\uparrow} \hat{n}_{\alpha\downarrow} + \sum_{\sigma\sigma',\alpha<\beta} (U' - J_H \delta_{\sigma\sigma'}) \hat{n}_{\alpha\sigma} \hat{n}_{\beta\sigma'} \nonumber \\
 &+ J_H \sum_{\alpha\neq\beta} (\hat{c}_{\alpha\uparrow}^\dagger \hat{c}_{\alpha\downarrow}^\dagger \hat{c}_{\beta\downarrow} \hat{c}_{\beta\uparrow} - \hat{c}_{\alpha\uparrow}^\dagger \hat{c}_{\alpha\downarrow} \hat{c}_{\beta\downarrow}^\dagger \hat{c}_{\beta\uparrow}) \nonumber \\
 &+ \Delta \sum_{\sigma} \hat{n}_{\sigma} + \frac{\lambda}{2} \hat{\bf c}^\dagger ({\bf L} \cdot {\bf s}) \hat{\bf c},
\end{align}
where $U$ is the intra-orbital (Hubbard) interaction, $U' = U - 2J_H$ the inter-orbital interaction, and $J_H$ the Hund's coupling. Further, $\lambda$ is the strength of the spin-orbit coupling (SOC) defined in terms of the vector of operators $\hat{\bf c} = (\hat{c}_{\alpha\uparrow}, \hat{c}_{\alpha\downarrow})$. The parameters of $H_U$ have been determined from first principles calculations, and are provided in Tab.~\ref{tab:parameters}.

For the Ni ions, the crystal field splitting is larger than the SOC, which further does not mix the
$e_g$ states~\cite{Stavropoulos2019}. In the half-filled $e_g$ manifold, the Hund's coupling therefore selects a spin triplet ground state with spin $S = 1$ and energy $U - 3J_H$. For the I ions, the spin-orbit coupling can be written in the basis $\hat{\bf c}^\dagger = (\hat{c}_{x\uparrow}^\dagger, \hat{c}_{x\downarrow}^\dagger, \hat{c}_{y\uparrow}^\dagger, \hat{c}_{y\downarrow}^\dagger, \hat{c}_{z\uparrow}^\dagger, \hat{c}_{z\downarrow}^\dagger)$, where the subscripts denote the $p$-orbital index, as
\begin{align}
 {\bf L} \cdot {\bf s} = \begin{pmatrix} 0 & -i\sigma_z & i\sigma_y \\
                                         i\sigma_z & 0 & -i\sigma_x \\
                                        -i\sigma_y & i\sigma_x & 0 \end{pmatrix}.
\end{align}
This splits the $p$-manifold into an effective $j = 1/2$ and $j = 3/2$ manifold, with energies $\lambda$ and $-\lambda/2$, respectively.

\subsection*{Hopping processes}
The magnetic interactions arise from the competition between local interactions and hopping processes between the Ni and I orbitals. In the following, we use $d_{i\alpha\sigma}$ and $p_{i\alpha\sigma}$ to denote the operators on the Ni and I ions, where $i$ denotes the ionic position, $\alpha$ the orbital type and $\sigma$ the local spin. To simplify the notation, we further introduce the short-hand labels $d_1 = d_{x^2-y^2}$ and $d_2 = d_{3z^2-r^2}$. 

Assuming inversion symmetry and $C_3$ symmetry around the $z$-axis, the magnetic interactions can be computed for a single Ni-Ni bond (see Figs.~\ref{fig:structure}c and \ref{fig:structure}d), and obtained for the remaining bonds by symmetry transformations~\cite{Maksimov2019}. Just like in honeycomb Kitaev systems~\cite{Winter2016}, the bonds can be divided into $X$-, $Y$- and $Z$-type bonds, where each bond involves hoppings along the complementary axes. The hopping processes on a $Z$-bond are~\cite{Stavropoulos2019}
\begin{align}
 H_{t1} &= \sum_{\langle ij\rangle\sigma} \Big(
    \Phi_{i\sigma}^\dagger {\bf M}^{(\gamma)} \Pi_{i+\gamma,\sigma}
  + \Phi_{j\sigma}^\dagger {\bf M}^{(-\gamma)} \Pi_{j-\gamma,\sigma} \Big) \\
  &+ {\rm H.c.} \nonumber
\end{align}
where $\Phi_{i\sigma}^\dagger = ( \hat{d}_{i,1,\sigma}^\dagger \; \hat{d}_{i,2,\sigma}^\dagger )$ and $\Pi_{i\sigma}^\dagger = ( \hat{p}_{ix\sigma}^\dagger \; \hat{p}_{iy\sigma}^\dagger \; \hat{p}_{iz\sigma}^\dagger )$ denote the Ni and I operators, $\gamma = -x$ and $+y$, and ${\bf M}^{(\gamma)}$ is a matrix describing the hopping between orbitals $\alpha$ and $\beta$ along crystal axis $\gamma$. Analogous expressions hold on the $X$ and $Y$ bonds. Assuming the symmetries mentioned above, the hopping parameters on different bonds are given by $t_1 = \sqrt{3}t/2$, $t_2 = t/2$ and $t_3 = t$. For Ni--I hoppings, the matrices along each crystal axis are given by
\begin{align}
 &M_{\alpha\beta}^{\pm x} = \pm \begin{pmatrix} t_1 & 0 & 0 \\ -t_2 & 0 & 0 \end{pmatrix} \quad M_{\alpha\beta}^{\pm y} = \pm \begin{pmatrix} 0 & -t_1 & 0 \\ 0 & -t_2 & 0 \end{pmatrix} \nonumber \\
 &M_{\alpha\beta}^{\pm z} = \pm \begin{pmatrix} 0 & 0 & 0 \\ 0 & 0 & t_3 \end{pmatrix}. \nonumber
\end{align}

There are also hopping paths involving direct Ni-Ni or I-I hopping, which were found to be sub-leading for the nearest neighbor interactions. The direct Ni-Ni hoppings are parameterized by
\begin{align}
 H_{t2} &= - \sum_{\langle ij\rangle\sigma} \Phi_{i\sigma}^\dagger \begin{pmatrix} r_1 & r_3 \\ r_3 & r_2 \end{pmatrix} \Phi_{j\sigma},
\end{align}
where $\Phi_{i\sigma}^\dagger = ( \hat{d}_{i,x^2-y^2,\sigma}^\dagger \; \hat{d}_{i,3z^2-r^2,\sigma}^\dagger )$.
while the direct I-I hoppings along a $Z$-bond are parameterized by
\begin{align}
 H_{t3} &= - \sum_{\langle ij\rangle\sigma} \Pi_{i\sigma}^\dagger 
 \begin{pmatrix} v_1 & v_2 & 0 \\ v_2 & v_1 & 0 \\ 0 & 0 & 0 \end{pmatrix} 
 \Pi_{j\sigma}. \nonumber
\end{align}
The full hopping Hamiltonian is given by the sum $H_t = H_{t1} + H_{t2} + H_{t3}$. All parameters entering $H_U$ and $H_t$ have been determined from first principles, and are given in Tab.~\ref{tab:parameters}.

\begin{table}
 \begin{tabular}{c|cccccccccc} \hline\hline
 Local & $U_{\rm Ni}$ & $U_{\rm I}$ & $J_{H,\rm Ni}$ & $J_{H,\rm I}$ & $\lambda_{\rm Ni}$ & $\lambda_{\rm I}$ & $\Delta$ & \T\B \\ \hline
 eV & 4.0 & 2.0 & 1.0 & 0.5 & 0.0 & 0.5 & -2.0 \T\B \\ \hline\hline
 Hoppings & $t_1$ & $t_2$ & $t_3$ & $v_1$ & $v_2$ & $r_1$ & $r_2$ & $r_3$ \T\B \\ \hline
 eV & 0.77 & 0.44 & 0.89 & 0.287 & 0.566 & 0.0 & 0.0 & 0.0 \T\B \\ \hline\hline
 Hoppings & $t_1$ & $t_2$ & $t_3$ & $v_1$ & $v_2$ & $r_1$ & $r_2$ & $r_3$ \T\B \\ \hline
 eV & 0.77 & 0.44 & 0.89 & 0.0 & 0.6 & 0.06 & 0.06 & 0.04 \T\B \\ \hline\hline 
 \end{tabular}
 \caption{{\bf Electronic parameters of NiI$_2$.} Parameters of the Hubbard-Kanamori Hamiltonian $H_U$ (top row), the hopping Hamiltonian $H_t$ for a nearest neighbor $Z$-bond (middle row, see Fig.~\ref{fig:clusters}a), and the hopping Hamiltonian $H_t$ for a third nearest neighbor $Z$-bond (bottom row, see Fig.~\ref{fig:clusters}b). The parameters were calculated from first principles using the {\sc Octopus}, {\sc Vasp} and {\sc Wannier90} electronic structure codes.}
 \label{tab:parameters}
\end{table}


\subsection*{Microscopic polarization operator}
To express the electric polarization operator in terms of spin operators, it is necessary to first represent it in the electronic basis. Similarly to the electronic hopping terms, the electric polarization operator can be written as
\begin{align}
 \hat{P}^a &= \sum_{\langle ij\rangle} \Big( \Phi_{i\sigma}^\dagger P_{\alpha\beta}^{a(\gamma)} \Pi_{j\sigma} + {\rm H.c.} \Big),
\end{align}
where $\gamma \in\{\pm x, \pm y, \pm z\}$ denotes the bond type, and $a$ gives the component of the polarization in the local crystal axes. On a $Z$-bond, all ions can be taken to lie in the local $xy$-plane, such that
\begin{align}
 \hat{P}_{ij}^a &= \Big( \hat{d}_{i\alpha\sigma}^\dagger P_{\alpha\beta}^{a(-x)} \hat{p}_{i-x\beta\sigma} + \hat{d}_{i\alpha\sigma}^\dagger P_{\alpha\beta}^{a(+y)} \hat{p}_{i+y\beta\sigma}  \\
 &\hspace*{0.2cm}+\hat{d}_{j\alpha\sigma}^\dagger P_{\alpha\beta}^{a(+x)} \hat{p}_{j+x\beta\sigma} + \hat{d}_{j\alpha\sigma}^\dagger P_{\alpha\beta}^{a(-y)} \hat{p}_{j-y\beta\sigma} \Big) + {\rm H.c.} \nonumber \\
 &P_{\alpha\beta}^{x(\pm x)} = \pm i\begin{pmatrix} 0 & t_1 & 0 \\ 0 & -t_2 & 0 \end{pmatrix} \quad P_{\alpha\beta}^{x(\pm y)} = 0 \nonumber \\
 &P_{\alpha\beta}^{y(\pm y)} = \pm i\begin{pmatrix} -t_1 & 0 & 0 \\ -t_2 & 0 & 0 \end{pmatrix} \quad P_{\alpha\beta}^{y(\pm x)} = 0 \nonumber
\end{align}
The polarization operator on $X$- and $Y$-bonds can be obtained by applying the symmetry operations of the systems.


\subsection*{Strong coupling expansion}
The effective spin-photon Hamiltonian is derived by dividing the electronic Hilbert space into a low- and a high-energy part, where the low-energy sector spans the local magnetic states $|m_i\rangle$. We define a projection operator $\mathcal{P}$ onto the low-energy sector, and a complimentary projector $\mathcal{Q} = 1 - \mathcal{P}$ onto the high-energy sector, both of which are multiplied by the unit operator $1$ in the photon Hilbert space. Since the magnetic states are eigenstates of the local Hamiltonian $H_U$, the coupling between the low- and high-energy sectors is mediated by the hopping Hamiltonian $H_t$. 

\begin{figure}
\includegraphics[width=\linewidth]{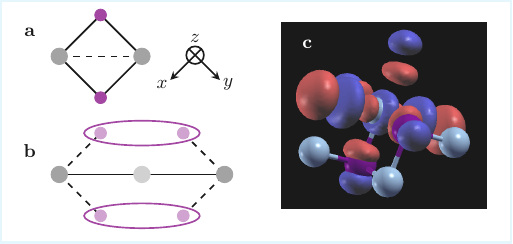}
 \caption{{\bf a,} Electronic cluster used to derive the nearest neighbor spin interactions. For the nearest neighbor interactions, the ligand mediated processes (solid lines) are found to give the dominant contribution, with sub-dominant contributions from direct (dashed lines) processes. {\bf b,} Electronic cluster used to derive the third nearest neighbor spin interactions, where the direct hopping processes (solid lines) are found to be dominant. To calculate $J_3$, we first diagonalize the effective I-I clusters, indicated by the purple ellipses, and only include the two highest energy states (separated from the filled lower states by an energy $3\lambda_I/2$) in the final calculation. {\bf c,} Wannier orbital centered on a Ni atom (light blue), calculated using {\sc vasp} and {\sc wannier90}. The Wannier functions are highly delocalized, and resemble molecular orbitals, explaining why the third nearest neighbor magnetic exchange is dominated by direct processes and of the same magnitude as the nearest neighbor exchange.}
 \label{fig:clusters}
\end{figure}

The coupling between low- and high-energy spaces can be eliminated order by order in $H_t$, by defining a unitary transformation $H = e^{-S} H e^{S}$ (called a Schrieffer-Wolff transformation) that block-diagonalizes the Hamiltonian $H$. Assuming that $S$ can be expanded in a small parameter $\gamma$, we can write $S = \gamma S_1 + \gamma^2 S_2 + \mathcal{O}(\gamma^3)$, to find the effective low-energy Hamiltonian
\begin{align}
 H_s &= H_0 + \gamma [S_1,V] + \frac{\gamma^2}{2} [S_1,[S_1,H_0]] + \mathcal{O}(\gamma^3).
\end{align}
The transformation matrices $S_i$ can be obtained in closed form, and can be found in Appendix B of Ref.~\cite{Winkler2003}. Once the transformation matrices have been constructed, it is a straightforward task to obtain the effective Hamiltonian. Here, we compute the spin-photon Hamiltonian by numerically implementing the Schrieffer-Wolff transformation up to fourth order in $\gamma$, on the cluster illustrated in Fig.~\ref{fig:clusters}a for nearest neighbor and in Fig.~\ref{fig:clusters}b for third nearest neighbor interactions.

For the polarization operator, we note that since $\hat{\bf P}$ has the form of a hopping Hamiltonian, it will be purely off-diagonal in the low- and high-energy subspace representation and $\mathcal{O}(\gamma)$ in the Schrieffer-Wolff expansion. Given the transformation matrices $S_i$ that block diagonalize the Hamiltonian, the low-energy representation of the polarization can be written as
\begin{align}
 \hat{\bf P}_s &= [\hat{\bf P}, \gamma S_1 + \gamma^2 S_2 + \gamma^3 S_3] + \frac{\gamma^3}{6} [\hat{\bf P}, S_1^3] \\
 &+ \frac{\gamma^3}{2} S_1 [S_1, \hat{\bf P}] S_1 + \mathcal{O}(\gamma^5). \nonumber
\end{align}
This follows from the fact that both $S_i$ and $\hat{\bf P}$ are purely off-diagonal, and so any non-zero contribution needs to be even in the number of such operators.

\subsection*{Effective single mode calculation}
For the numerical results discussed in the main text, we use the single effective mode approximation together with a fourth order Schrieffer-Wolff transformation, including all hopping processes on a single $Z$-bond cluster. Here, we discuss the analytical result for the simplified model introduced in the Supplemental Material, which is equivalent to a single-band Hubbard model with effective hopping $t_i$ ($i = 1,3$) and interacion $U + J_H$. 

In the single mode limit, the expression for the magnetic exchange simplifies to
\begin{align}\label{eq:single_mode}
 \frac{J}{J_{0}} &= e^{-|g|^2} \sum_n \frac{|g|^{2n}}{n!} \frac{\Omega}{\Omega + \omega_s n}
\end{align}
where $\Omega = (U+J_H)/\hbar$ and $n$ runs over the occupation of the number state $|n\rangle$. We note that in the limit $\Omega \to \infty$, the second term of the sum becomes unity and it is straightforward to perform the sum. However, this exactly cancels the exponential prefactor, such that the effect of the cavity is zero. For finite $\Omega$, we define $\beta = \omega_s/\Omega$ and use the expression $(1 + a)^{-1} = \int_0^\infty d\alpha\, e^{-(1 + a) \alpha}$ to write the sum as an integral. After some manipulations we find
\begin{align}
 \frac{J_i}{J_{i,0}} &= \frac{e^{-|g|^2}(-|g|^2)^{-1/\beta}}{\beta} \Big[ \Gamma \Big(\frac{1}{\beta} \Big) - \Gamma \Big(\frac{1}{\beta}, -|g|^2 \Big) \Big],
\end{align}
where $\Gamma(s,x)$ is the upper incomplete $\Gamma$-function.

The above expression can be simplified by noting that $\Gamma(a,-|g|^2) = \Gamma(a) - \gamma(a,-|g|^2)$, where $\gamma$ is the lower incomplete $\Gamma$-function satisfying $\gamma(a,-|g|^2) \sim (-|g|^2)^a/a$ for small $g$. This gives
\begin{align}
 \frac{J_i}{J_{i,0}} &\sim \frac{e^{-|g|^2}(-|g|^2)^{-1/\beta}}{\beta} (-|g|^2)^{1/\beta} \beta = e^{-|g|^2}.
\end{align}
This result can also be obtained from Eq.~\ref{eq:single_mode} by taking the leading order term in $g$, and can therefore be viewed as an expansion in the number of photons involved in the virtual state. This expansion, continued to next-to-leading order, gives the form $X/X_0 = \exp(-\alpha|g|^2) (1 + \beta |g|^2)$ of the cavity modification used to fit the numerical data in the main text.


\subsection*{Comparison to the multi-mode calculation}\label{app:multi_mode}
For the simplified model discussed above, the effect of the SPP fluctuations can be worked out analytically including the full SPP mode structure. The full calculation is presented in Ref.~\cite{GrunwaldCheng}, and results in the expression
\begin{align}\label{eq:exact_J}
 \frac{J}{J_0} &= e^{-\alpha} \int_0^\infty dx\, \exp \Big[ \alpha e^{-\beta x} - x \Big] \\
 &= \frac{e^{-\alpha}(-\alpha)^{-1/\beta}}{\beta} \Big[ \Gamma \Big(\frac{1}{\beta} \Big) - \Gamma \Big(\frac{1}{\beta}, -\alpha \Big) \Big]. \nonumber
\end{align}
Identifying $\alpha = |g|^2$, this is in exact agreement with the single mode expression above. Here, we show that this choice of $\alpha$ arises naturally from matching the exact expression to the leading order result in the light-matter coupling.

We note that to leading order in $\alpha$, Eq.~\ref{eq:exact_J} is $J/J_0 \approx 1 - \alpha$. However, to leading order in the light-matter couplings $g_\bq$, we can also write the cavity modification as
\begin{align}
 \frac{J}{J_0} &= 1 - \sum_{\bq} g_{\bq}^2 \frac{\Omega}{\Omega + \hbar\omega_s} \\
 &= 1 - \frac{e^2a^2\Omega(\omega_s^2 - \omega_{\rm TO}^2)}{4\hbar(\Omega + \omega_s) \epsilon_0\epsilon_r \omega_s^3} \frac{1}{A} \sum_\bq  qe^{-2qd}. \nonumber
\end{align}
Here $\epsilon_r = \epsilon_{\rm sub} + \epsilon_{\rm mat}$, $J_0$ is the exchange coupling as the substrate-material distance approaches infinity, and $\Omega = (U + J_H)/\hbar$ is the characteristic electronic frequency. The sum can be evaluated in the macroscopic limit by converting it to an integral, such that the modified exchange interaction is
\begin{align}
 \frac{J_1}{J_0} - 1 &= -\frac{e^2a^2\Omega(\omega_s^2 - \omega_{\rm TO}^2)}{16\pi\hbar d^3 (\Omega + \omega_s) \epsilon_0 \epsilon_r \omega_s^3} \\
 &\approx -\frac{e^2a^2}{(\hbar\omega_s)^2} \frac{\hbar^2(\omega_s^2 - \omega_{\rm TO}^2)}{16 \pi\hbar\omega_s \epsilon_0 \epsilon_r d^3} = |g|^2. \nonumber
\end{align}
By choosing $\alpha = |g|^2$, the single mode and multi-mode expressions are in agreement to all orders in the light-matter coupling.


\subsection*{Details on the first principles calculations}
To obtain the single-particle parameters of NiI$_2$, we performed density functional theory (DFT) calculations using the projector-augmented wave method, as implemented in the {\sc vasp} electronic structure code. The Ni $3d$ and $4s$ orbitals, as well as the I $5s$ and $5p$ orbitals, were explicitly included in the calculation, with a plane-wave cut-off of $350$ eV. We used the Perdew-Burke-Erzenhof (PBE) exchange-correlation functional and performed PBE$+U$ calculation within the Dudarev approach, with $U = 4$ eV and the spin-orbit coupling fully considered. To constrain the spins into a given pattern, we used the penalty functional implemented in {\sc vasp} with $\hbar\omega = 1$ eV. To accomodate the spiral order we used a $7 \times 1 \times 1$ real-space supercell, and a $k$-point mesh of $1 \times 6 \times 1$.

The Wannierization was performed with {\sc wannier90}, using the electronic structure calculated from {\sc vasp}. This was done by first separately Wannierizing two copies of the system with opposite magnetization, one along the positive $c$-direction, and the other along the negative $c$-direction. As the two copies are time-reversal partners of each other, invariance under time reversal for the non-magnetic phase was ensured by averaging the resulting tight-binding models over the two Wannier copies.

To determine the parameters of the local electronic Hamiltonian $H_U$, we performed similar calculations with the {\sc octopus} electronic structure code. The interaction parameters where determined using the hybrid DFT$+U$ functional ACBN0 in the ferromagnetic state. Mixed boundary conditions, periodic in the in-plane direction and open in the out-of-plane direction, where used together with a vacuum region of $20$ {\AA} to ensure convergence in the out-of-plane direction. A $12 \times 12$ $k$-point grid and a real-space grid spacing of $0.3$ Bohr were employed. Using the ACBN0 functional, a self-consistent effective interaction $U_{\rm eff} = U - J_H$ was determined on Ni ions, and the Kanamori parameters $U$ and $J_H$ where calculated in the final state after convergence had been reached.


\subsection*{Details on the Monte Carlo simulation}
Classical spin Monte Carlo simulations were performed on a triangular lattice spanned by the primitive vectors $a_1 \parallel [100]$ and $a_2 \parallel [110]$, using a system with $N_s = 80 \times 80$ sites  and periodic boundary condition. The temperature was annealed from $ T = 68.5$ K to $ 0 $ K, in steps of $ 1.16 $ K. The initial spin configurations were generated randomly: each spin was assigned an isotropically distributed vector on the unit sphere. We employed a Metropolis scheme with single spin trial moves. One Monte Carlo sweep contains $N_s$ attempts of spin flips, and measurements were separated by $20$ sweeps to reduce autocorrelation. We performed a pre‑thermalization of $200$ sweeps at the initial temperature, and then generated $300$ measurements per temperature, discarding the first $150$ as equilibration and using the remaining $150$ for statistics. 

For each measurement, we calculated the discrete Fourier transform of the spin configuration and the corresponding polarization configuration. From the Fourier spectrum, we extracted the maximum $\max_\bq |{\bf S_{q}}|$ of the spin structure, its location $\bf q$ (determining the spin order), and the squared magnitude of the polarization $|{\bf P}|^2$. For each parameter set, $128$ independent simulations with distinct random seeds were performed. Equilibration was checked using the total energy, and the chosen measurement spacing was found to yield negligible correlation between each measurement. Finite size effects were evaluated by repeating simulations for $N_s = 60 \times 60$ and $N_s = 100 \times 100$. Observables reported agree within statistical uncertainty, indicating that $N_s = 80 \times 80$ is sufficient.


\subsection*{Topological charge on a lattice}
The topological charge on the triangular lattice spin configuration is defined as~\cite{Berg81,vinas_bostrom_microscopic_2022}
\begin{align}
    Q = \frac{1}{4\pi} \sum_{ijk} \Omega_{ijk},
\end{align}
where $\Omega_{ijk}$ is defined as the solid angle on the sphere spanned by three neighboring spins. The explicit definition is
\begin{align}
    \exp\Big( \frac{i\Omega_{ijk}}{2} \Big) &= \frac{1}{\rho} \Big[ 1 + {\bf S}_i\cdot{\bf S}_j + {\bf S}_j\cdot{\bf S}_k + {\bf S}_k\cdot {\bf S}_i \\ 
    &+ i {\bf S}_i\cdot({\bf S}_j\times{\bf S}_k) \Big], \nonumber
\end{align}
where $\rho = \sqrt{2(1+{\bf S}_i\cdot{\bf S}_j)(1+{\bf S}_j\cdot{\bf S}_k)(1+{\bf S}_k\cdot {\bf S}_i)}$. Here we always take the $i\to j\to k\to i$ path to be counterclockwise. We calculate the charge density on the lattice and finding that merons (antimerons) form in pairs, resulting in total integer charge $Q$ on the lattice if the system is in the transition region between the FM and HM phases. In the HM phase, meron and antimeron are bound together in chains along a domain wall, giving a total charge $Q = 0$.


\bibliography{references}


\clearpage


\setcounter{equation}{0}
\renewcommand{\theequation}{S\arabic{equation}}

\setcounter{figure}{0}
\renewcommand{\thefigure}{S\arabic{figure}}

\section*{Supplemental Material}


\subsection*{Analytic derivation of the spin model}
To supplement the numerical results, we also derive an analytical expression for the spin parameters $J_1$ and $J_3$ using a simplified model. We assume the hopping processes mediated by the ligands can be subsumed into an effective Ni-Ni hopping, thereby defining the effective hopping matrix
\begin{align}
 T_{ij} = \begin{pmatrix} t_1 & t_3/\sqrt{2} \\ t_3/\sqrt{2} & t_2 \end{pmatrix}.
\end{align}
To second order in the hopping, the effective spin Hamiltonian is then given by
\begin{align}\label{eq:ham_eff}
 H_s &= \sum_{\langle ij\rangle} \mathcal{P} \Phi_{i\sigma}^\dagger T_{ij} \Phi_{j\sigma} \mathcal{Q} \bigg[ \frac{1}{E_0 - H_0} \bigg] \mathcal{Q} \Phi_{j\sigma'}^\dagger T_{ji} \Phi_{i\sigma'} \mathcal{P}.
\end{align}
Here $\mathcal{P}$ is the projection operator onto the low-energy $S = 1$ manifold, $\mathcal{Q} = 1 - \mathcal{P}$ is the projector onto the complement, $E_0 = 2E_t = 2U - 6J$ is the energy of the Ni triplet state, and $H_0 = H_U$.

To derive the spin Hamiltonian, we evaluate the matrix elements of Eq.~\ref{eq:ham_eff} in the local spin eigenstates $|m_1 m_2\rangle = |S_1 m_1, S_2 m_2\rangle$. The labels $S_1$ and $S_2$ can be suppressed since we always have $S_1 = S_2 = 1$. Starting from a configuration with two holes on each Ni, the intermediate state will always be a product of a one-hole state and a three-hole state, and have the energy $E = E_1 + E_3 = 3U - 5J$. We therefore have $E_0 - E = - U - J$, and within the triplet subspace we can write
\begin{align}
 H_s &= \frac{-1}{U + J} \sum_{\langle ij\rangle} \mathcal{P} \Phi_{i\sigma}^\dagger T_{ij} \Phi_{j\sigma} \mathcal{Q} \mathcal{Q} \Phi_{j\sigma'}^\dagger T_{ji} \Phi_{i\sigma'}.
\end{align}
To work out the matrix elements, we note that the Hamiltonian has the structure $H_{s,ij} = M_{ji}^\dagger M_{ji} + M_{ij}^\dagger M_{ij}$, where $M_{ji}$ is a $9 \times 16$ matrix that takes a given triplet state $|m_1 m_2\rangle$ into state expressible in the product basis $|a_m b_n\rangle$, where $|a_m\rangle$ is a one-hole state and $|b_n\rangle$ is a three-hole state. Evaluating these matrices we find an effective spin Hamiltonian
\begin{align}
 H_s &= J \sum_{\langle ij\rangle} {\bf S}_i \cdot {\bf S}_j.
\end{align}
For the simplified model considered here, we find with $t_{\rm eff}^2 = -2(t_1^2 + t_2^2 + t_3^2)$ that the effective exchange coupling is
\begin{align}
 J &= \frac{t_{\rm eff}^2}{U+J}.
\end{align}


\subsection*{Cavity light-matter coupling}
We now consider the effect of electromagnetic vacuum fluctuations on the equilibrium magnetic state of NiI$_2$. In presence of an electromagnetic field, the total Hamiltonian is given by $H = H_U + H_t + \sum_{\lambda} \hbar\Omega_\lambda \hat{n}_\lambda$, where $\hbar\Omega_\lambda$ is the energy of cavity mode $\lambda$, and $\hat{n}_\lambda$ is the corresponding number operator. Because of fluctuations of the electromagnetic field, the hopping amplitudes acquire an additional phase, and the Hamiltonian $H_t$ is modified by the replacement
\begin{align}
 \hat{c}_{i\alpha\sigma}^\dagger \hat{c}_{i\beta\sigma} \to e^{i \phi_{ij}} \hat{c}_{i\alpha\sigma}^\dagger \hat{c}_{i\beta\sigma}.
\end{align}
Here $\phi_{ij} = (ea/\hbar)\, {\bf r}_{ij} \cdot \hat{\bf A}$ is a Peierls phase, ${\bf r}_{ij} = {\bf r}_j - {\bf r}_i$ is the vector between atomic sites $i$ and $j$ (measured in units of the Ni-Ni distance $a$), and the quantum vector potential is
\begin{align}
 \hat{\bf A} = \sum_{\lambda} ( A_\lambda {\bf e}_\lambda \hat{a}_\lambda^\dagger + A_\lambda^{*} {\bf e}^{*}_\lambda \hat{a}_\lambda).
\end{align}
Here $A_\lambda$ is a mode function, and for a given mode $\lambda$ this scheme defines the dimensionless light-matter coupling $g_\lambda = (ea/\hbar) A_\lambda = ea/\sqrt{2\epsilon_0 \hbar\Omega_\lambda V}$. The Peierls phases can be written as in terms of dimensionless variables as $\phi_{ij} = {\bf r}_{ij} \cdot \hat{\bf a}$ with
\begin{align}
 \hat{\bf a} = \sum_{\lambda} ( g_\lambda {\bf e}_\lambda \hat{a}_\lambda^\dagger + g_\lambda^{*} {\bf e}^{*}_\lambda \hat{a}_\lambda),
\end{align}
and we note that $\phi_{ij}$ only depends on the positions of the ions involved. The modified hopping Hamiltonian and polarization operator is obtained by adding the appropriate Peierls phases to each of the bonds.

The total Hamiltonian can be expanded in the photon number basis $|{\bf n}\rangle = |n_1, n_2, \ldots, n_N \rangle$ according to~\cite{VinasBostrom2023}
\begin{align}
 H &= \sum_{\bf nm} ({\bf 1}_e \otimes |{\bf n}\rangle \langle {\bf n}|) H ({\bf 1}_e \otimes |{\bf m}\rangle \langle {\bf m}|) \nonumber \\
 &= \sum_{\bf nm} H_{\bf nm} \otimes |{\bf n}\rangle \langle {\bf m}|,
\end{align}
where ${\bf 1}_e$ is the identity operator in the electronic Hilbert space, and the Hamiltonian $H_{\bf nm}$ is given by
\begin{align}
 H_{\bf nm} &= \big( H_U + \sum_\lambda \hbar\Omega_\lambda n_\lambda \big) \delta_{\bf nm} + H_{t,\bf nm}.
\end{align}
To calculate the matrix elements $\langle {\bf n}| e^{i{\bf d}_{ij} \cdot \hat{\bf a}} |{\bf m}\rangle$ we note that since $[\hat{a}_\lambda^\dagger, \hat{a}_{\lambda'}] = 0$, the Peierls phases factorize over different modes and
\begin{align}
 \langle {\bf n}| e^{i{\bf d}_{ij} \cdot \hat{\bf a}} |{\bf m}\rangle = \prod_\lambda \langle n_\lambda| e^{i{\bf d}_{ij} \cdot \hat{\bf a}_\lambda} |m_\lambda\rangle.
\end{align}
The single-mode expectation values are calculated by introducing the variables $\eta_{ij\lambda} = g_\lambda ({\bf d}_{ij} \cdot {\bf e}_\lambda)$, and using the Baker-Hausdorff expansion for the exponential. The matrix elements are $\langle n_\lambda| e^{i{\bf d}_{ij} \cdot \hat{\bf A}_\lambda} |m_\lambda\rangle = i^{|n_\lambda - m_\lambda|} j_{n_\lambda,m_\lambda}^{ij}$, where explicit forms of $j_{n_\lambda,m_\lambda}^{ij}$ are provided in Ref.~\cite{VinasBostrom2023}. Using the notation ${\bf g} = \{ g_1, g_2, \ldots, g_N \}$ to denote the set of couplings, the light-matter interaction is described by the function
\begin{align}
 J_{\bf nm}^{ij}({\bf g}) = \langle {\bf n}| e^{i{\bf d}_{ij} \cdot \hat{\bf A}} |{\bf m}\rangle = \prod_\lambda j_{n_\lambda,m_\lambda}^{ij}.
\end{align}
With this function, the hopping Hamiltonian in presence of the cavity can be written as
\begin{align}\label{meth:hopping_photon}
 H_{t2,\bf nm} &= -\sum_{\langle ij\rangle\sigma} \Phi_{i\sigma}^\dagger J_{\bf nm}^{ij}({\bf g})
 \begin{pmatrix} r_1 & r_3 \\ r_3 & r_2 \end{pmatrix}
 \Phi_{j\sigma},
\end{align}
here exemplified for a Ni-Ni bond.


\subsection*{Surface phonon polaritons}\label{app:spp}
The discussion above holds for any cavity mode structure, but we now specialize to a surface cavity where the electric field comes from surface phonon polaritons (SPP) of a paraelectric surface~\cite{VinasBostrom2024}. The electric field corresponding to the SPPs can be written as
\begin{align}
 {\bf E}_\parallel({\bf x},t) &= -i \sum_\bq f_q \frac{{\bf q}}{|q|} e^{-qd} e^{i {\bf q} \cdot {\bf x} - i \omega_s t}  \left( \hat{a}_\bq^\dagger + \hat{a}_{-\bq} \right) \\
  E_{\perp}({\bf x},t) &= \sum_\bq f_q e^{-qd}e^{i {\bf q} \cdot {\bf x} - i \omega_s t}  \left( a^\dag_\bq + a_{-\bq} \right) \nonumber \\
 f_q &= \sqrt{\frac{q \hbar (\omega_s^2 - \omega_{\rm TO}^2)}{4 A \epsilon_0 (\epsilon_{\rm sub} + \epsilon_{\rm mat}) \omega_s}}, \nonumber
\end{align}
where $\omega_s^2 = (1/2) (\omega_{\rm LO}^2 + \omega_{\rm TO}^2)$, and $\omega_{\rm LO}$ and $\omega_{\rm TO}$ are the longitudinal and transverse optical phonon frequencies of the substrate. For these modes, we can calculate the local electric field noise $\langle {\bf E}^2\rangle$, which has the form~\cite{VinasBostrom2024}
\begin{align}
 \epsilon_0 \langle {\bf E}^2 \rangle &= \frac{\hbar^2(\omega_s^2 -\omega_{\rm TO}^2)}{16 \pi\hbar\omega_s \epsilon_r d^3}.
\end{align}
As discussed in the main text, this can be used to construct a single effective mode approximation.

The SPP electric field can be compared with that at the center of a Fabry-Perot cavity, where the local electric field noise is $\epsilon_0 \langle {\bf E}^2 \rangle = 7\pi^2\omega_c/1080h^3$~\cite{VinasBostrom2024}. Here $\omega_c = c\pi/h$ is the fundamental cavity frequency and $h$ is the height of the cavity, which for a Fabry-Perot cavity are not independent. For a THz cavity we find $h \sim 10$ $\mu$m, while $d$ is independent of frequency and can be taken as $d \sim 10$ nm, this leads to a large enhancement $\eta = \epsilon_0 \langle {\bf E}^2 \rangle_{\rm SPP}/\epsilon_0 \langle {\bf E}^2 \rangle_{\rm FP} \sim h^3/d^3 \sim 10^9$ of the effective electric field.


\subsection*{Analytic solution for the $J_1-J_3-A_{zz}$ model}
To analytically estimate the ground state energy of the helical state in the macroscopic limit, we consider an effective $J_1-J_3-A_{zz}$ model on a triangular lattice. This is expected to provide a good approximation to the true ground state, since the full spin Hamiltonian is dominated by the competition between the $J_1$ and $J_3$ terms. We compare the energies of two helical magnetic states, whose propagation vectors are along the $[1\bar10]$ and $[100]$ directions, and the ferromagnetic state.

For each classical spin ${\bf S}_i$, with length $|{\bf S}_i| = 1$, the next spin along the propagation direction is obtained by the rotation ${\bf S}_{i+1} = R(\theta,\varphi,\alpha) {\bf S}_i$, while the previous one is ${\bf S}_{i-1} = R(\theta,\varphi,-\alpha) {\bf S}_i$.
The rotation matrix $R(\theta,\varphi,\alpha)$ is defined by the polar and azimuthal angles $\theta$ and $\varphi$ of the normal vector of the spin rotational plane, and the relative angle $\alpha \in [0,\pi]$ within this plane. If $\alpha = 0$, the system is in the ferromagnetic phase. With a positive single-ion anisotropy $A_{zz}$, a finite out-of-plane component $S_z$ will increase the energy of the magnetic state, and so the ground state is described by an effective $XY$ model with $\theta = 0$. The Hamiltonian of the single helix states with $\bq_1 \parallel [1\bar{1}0]$ and $\bq_2 \parallel [100]$ are then
\begin{align}
    H_{q_1} &= N \Big[ J_1 (1 + 2\cos\alpha) + J_3 \left(1 + 2\cos2\alpha\right) \Big], \\
    H_{q_2} &= N \Big[ J_1 (2\cos\frac{\alpha}{2} + \cos\alpha) + J_3 (2\cos\alpha + \cos 2\alpha) \Big].
\end{align}
In NiI$_2$ the nearest neighbor interaction is ferromagnetic, $J_1 < 0$, while the third nearest neighbor is antiferromagnetic, $J_3 > 0$, and we denote their ratio by $\eta = |J_1|/J_3 > 0$. To ease the notation we write $x_1 = \cos \alpha \in [-1,1]$ and $x_2 = \cos (\alpha/2) \in [-1,1]$, such that the Hamiltonian is
\begin{align}
    H_{q_1} &= N J_3 (4x_1^2 - 2\eta x_1 - \eta - 1), \label{eq:Hq1} \\
    H_{q_2} &= N J_3 \left[ 8x_2^4 - (2\eta + 4) x_2^2 - 2\eta x_2 + \eta - 1 \right]. \label{eq:Hq2}
\end{align}
For Eq.~\ref{eq:Hq1}, the minimum occurs at
\begin{equation}
    \begin{aligned}
        x_1 = \left\{
        \begin{array}{l}
         \frac{\eta}{4}, \quad \eta\in(0,4]\\
         1, \quad \eta\in(4,+\infty)
        \end{array}  \right.
    \end{aligned}
\end{equation}
For Eq.~\ref{eq:Hq2}, the minimum occurs at
\begin{equation}
    \begin{aligned}
        x_2 =\left\{\begin{array}{l}
        \frac{1}{4}(1+\sqrt{1+2\eta}), \quad \eta\in(0,4]\\
        1, \quad\quad\quad\quad\quad\quad\quad\eta\in(4,+\infty)
        \end{array}  \right.
    \end{aligned}
\end{equation}
In both cases, the system transitions from a ferromagnetic to a helimagnetic state at $\eta = 4$, or $|J_1| = 4J_3$. We can also determine the more stable helical phase by comparing the ground state energies $E_{0,q_i}$ of the Hamiltonian $H_{q_1}$ and $H_{q_2}$. Their ground state energies are
\begin{align}
    {E}_{0,q_1} &= NJ_3 \bigg( -\eta - 1 - \frac{\eta^2}{4} \bigg), \\
    {E}_{0,q_2} &= \frac{NJ_3}{8} (-4\eta\sqrt{1 + 2\eta} + 2\eta - 2\sqrt{1+2\eta} - 10 - \eta^2).
\end{align}
We find that $E_{0,q_2} \leq  E_{0,q_1}$ for all $\eta \in(0,4]$, showing that the helix order prefers to propagate along $[100]$ direction.

\begin{figure}[h]
    \centering
    \includegraphics[width=0.9\linewidth]{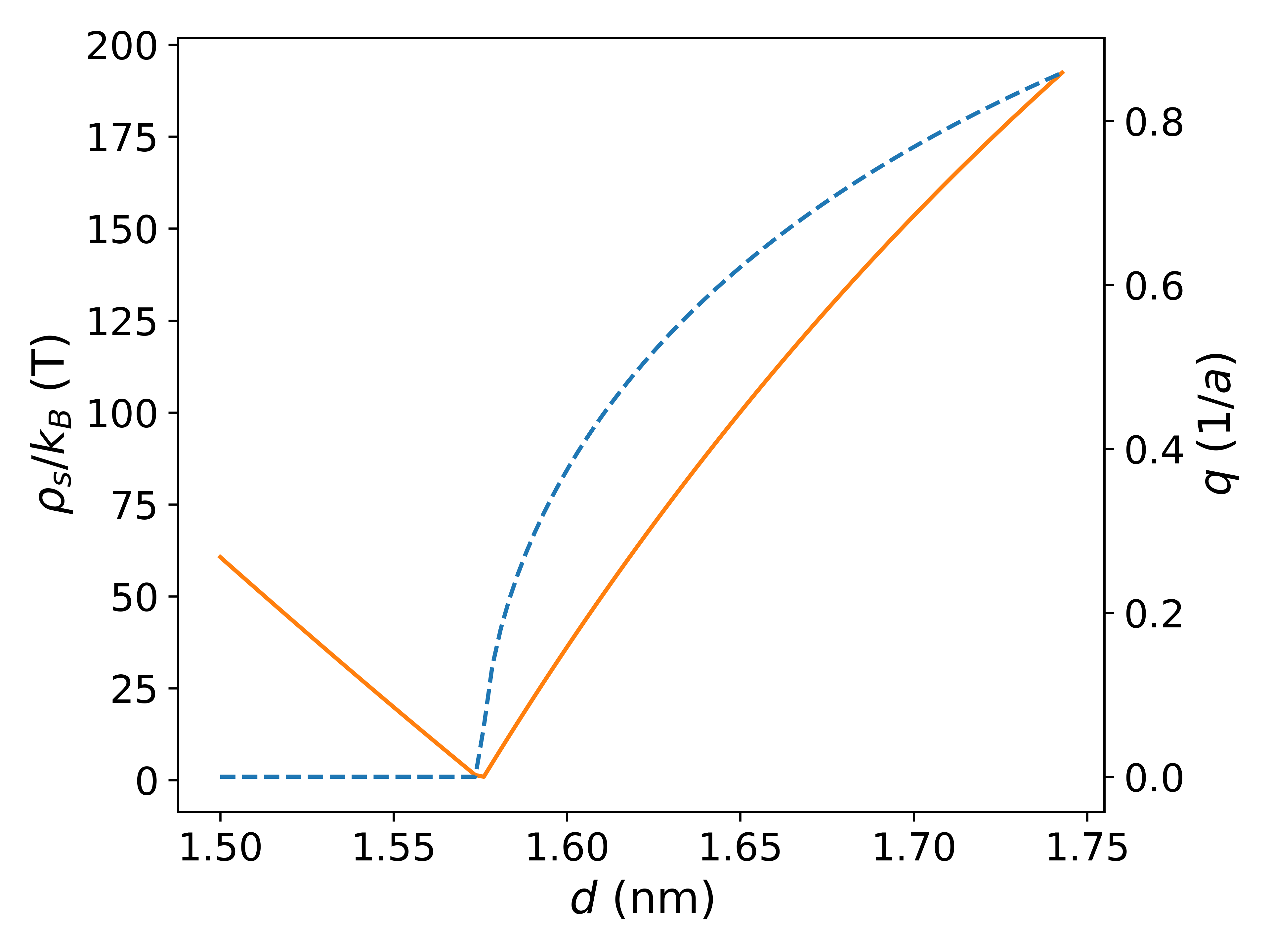}
    \caption{Theoretically calculated spiral momentum $q$ (blue dashed line), related to $\alpha_0$, and the classical spin stiffness $\rho_s/k_B$ (orange solid line) of NiI$_2$.}
    \label{fig:stiffness}
\end{figure}

Furthermore, we also evaluated the spin stiffness for the preferred helix state, which is defined as
\begin{align}
    \rho_s &= \frac{1}{N}\frac{\partial^2 H_{q_2}}{\partial \alpha^2}\Big|_{\alpha=\alpha_0}, \\
    & = -2J_1 \cos\frac{\alpha_0}{2} - (4J_1 + 8J_3)\cos\alpha_0 -16 J_3 \cos2\alpha_0. \nonumber
\end{align}
In Fig.~\ref{fig:stiffness}, we show the theoretically determined $\alpha_0$ and the corresponding classical spin stiffness $\rho_s/k_B$ as functions of the distance $d$ to the surface cavity. Near the critical distance $d_c \approx 1.6$ nm, the stiffness approaches zero, indicating a softening of the stretching mode and an enhanced sensitivity to quantum fluctuations in this regime. A more comprehensive understanding of the physics in the vicinity of $d_c$ would require going beyond a classical description and thermal fluctuations, and to analyze the corresponding quantum spin model on the triangular lattice.

\subsection*{Specific heat}
\begin{figure}[h]
    \centering
    \begin{tikzpicture}
     \node at (0, 3.0) {\includegraphics[width=0.9\linewidth]{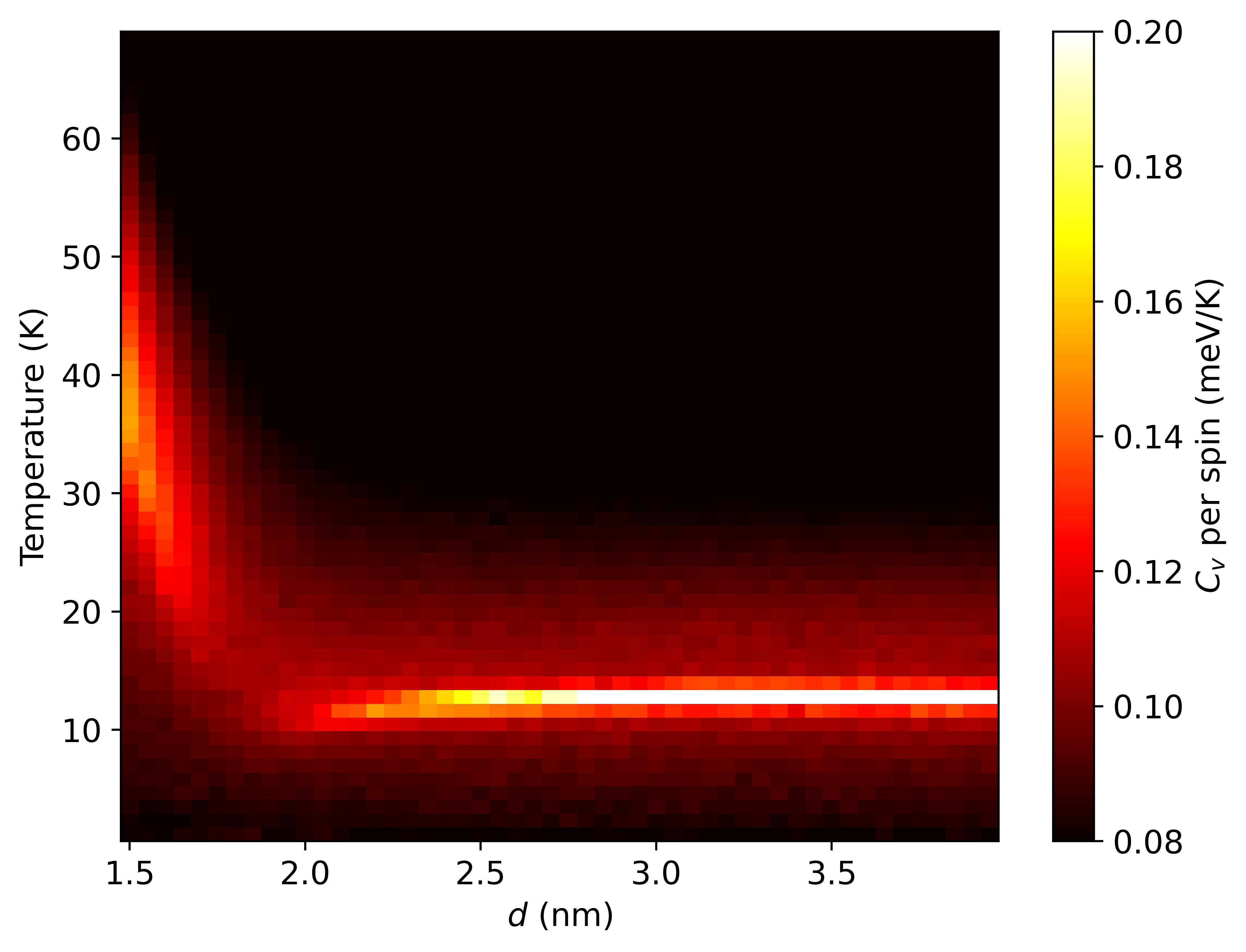}};
     \node at (0,-3.0) {\includegraphics[width=0.9\linewidth]{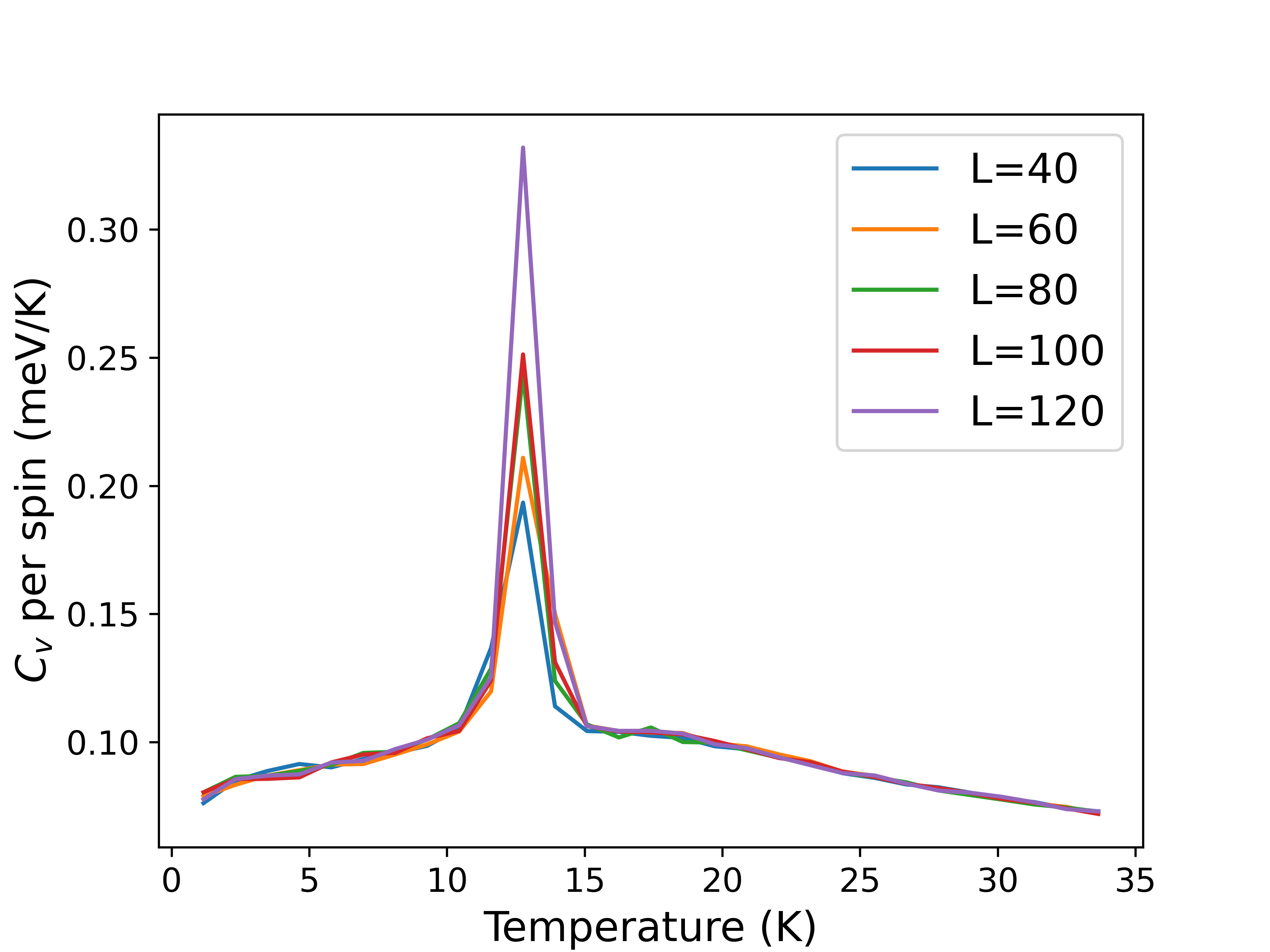}};
     \node[white] at (-2.4, 5.1) {\bf a};
     \node[black] at (-2.3,-1.4) {\bf b};
    \end{tikzpicture}
    \caption{{\bf a,} Simulated specific heat per spin, $C_v$, as a function of distance $d$ and temperature $T$. {\bf b,} Simulated specific heat per spin, $C_v$, as a function of temperature $T$ for different system sizes $L$. The distance is chosen as $d = 3.95$ nm.}
    \label{fig:Cv_finite_size_scaling}
\end{figure}

From the Monte Carlo simulations, we obtain the specific heat per spin $C_v(T)$ as a function of distance $d$ and temperature $T$. It is defined as
\begin{equation}
    C_v(T) = \frac{\braket{E^2}-\braket{E}^2}{Nk_BT^2},
\end{equation}
where $E$ is the measured total energy. The results in Fig.~\ref{fig:Cv_finite_size_scaling} show a large specific heat at large $d$, for temperature $T \approx 12.8$ K, and a finite size scaling consistent with a second order phase transition from a paramagnetic to a helimagnetic phase. Compared with Fig.~\ref{fig:numerics}c of the main text, the phase boundaries are similar, apart from the low temperature transition from the HM and FM phases. From an energy perspective, these transitions show a smooth crossover rather than a second order phase transition. Together with the spin stiffness, these findings show that topological defects enrich the nature of these transitions, making the meron phase worthy of further study.

\end{document}